%% file: 0.main_paper.tex
\documentclass[conference]{IEEEtran}
\usepackage{cite}
\usepackage{amsmath,amssymb,amsfonts}
\usepackage{algorithmic}
\usepackage{graphicx}
\usepackage{textcomp}
\usepackage{xcolor}
\usepackage{subcaption}
\usepackage{adjustbox,booktabs}
\usepackage{multirow}
\usepackage{caption}
\usepackage{tabularx}
\usepackage{diagbox}
\usepackage{hhline}
\usepackage{arydshln}
\usepackage{cellspace}
\def\BibTeX{{\rm B\kern-.05em{\sc i\kern-.025em b}\kern-.08em
    T\kern-.1667em\lower.7ex\hbox{E}\kern-.125emX}}
\begin{document}

\title{Sleep Model -- A Sequence Model for Predicting the Next Sleep Stage
}

\author{\IEEEauthorblockN{Iksoo Choi and Wonyong Sung}
	\IEEEauthorblockA{\textit{Department of Electrical and Computer Engineering} \\
	\textit{Seoul National University} Seoul, Korea \\
	\texttt{\{akacis, wysung\}@snu.ac.kr}}
}

\maketitle

\begin{abstract}
	As sleep disorders are becoming more prevalent there is an urgent need to classify sleep stages in a less disturbing way.
	In particular, sleep-stage classification using simple sensors, such as single-channel electroencephalography (EEG), electrooculography (EOG), electromyography (EMG), or electrocardiography (ECG) has gained substantial interest.
	In this study, we proposed a sleep model that predicts the next sleep stage and used it to improve sleep classification accuracy. 
	The sleep models were built using sleep-sequence data and employed either statistical $n$-gram or deep neural network-based models. 
	We developed beam-search decoding to combine the information from the sensor and the sleep models. 
	Furthermore, we evaluated the performance of the $n$-gram and long short-term memory (LSTM) recurrent neural network (RNN)-based sleep models and demonstrated the improvement of sleep-stage classification using an EOG sensor.
	The developed sleep models significantly improved the accuracy of sleep-stage classification, particularly in the absence of an EEG sensor.
\end{abstract}


\input{1.Introduction.tex}
\input{2.RelatedWorks.tex}
\input{3.SleepModel_BeamSearchDecoding.tex}
\input{4.ExperimentalResults.tex}
\input{5.ConculdingRemark.tex}

\bibliographystyle{IEEEbib}
\bibliography{a_.References}

\end{document}

%% file: 1.Introduction.tex
\section{Introduction}\label{sec:intro}
Currently, sleep disorders are prevalent and lead many people to seek help from a psychiatrist or specialist.
The sleep test divides a sleep period into epochs, typically 30-second long, and assigns a sleep class label to each epoch.
The classification guidelines of the American Association of Sleep Medicine (AASM) distinguish five sleep classes: rapid eye movement ($REM$) sleep, three non-$REM$ sleep classes ($N1$, $N2$, and $N3$), and wake ($W$)~\cite{berry2017aasm}.
Sleep staging is generally performed through manual visual scoring of polysomnography (PSG) that includes several signal sources, such as electroencephalography (EEG), electrooculography (EOG), electromyography (EMG), electrocardiography (ECG), and respiratory sensors!\cite{kushida2005practice}.

While PSG remains the gold standard for the clinical evaluation of sleep, it is impractical for long-term monitoring of sleep problems at home. 
In recent years, several surrogate measures have been studied to reduce the costs and discomfort associated with PSG testing. 
Many sleep classification studies have employed a subset of PSG sensors, including single-channel EEG, EMG, EOG, ECG, or PPG~\cite{perslev2019u,rahman2018sleep,radha2021deep,li2018deep,fonseca2015sleep,choi2022performance}. 
However, when compared to full PSG, these surrogate tests inevitably result in an accuracy drop. Deep neural networks (DNN) offer new opportunities to dramatically improve the accuracy of simple sleep tests~\cite{radha2019sleep,phan2019seqsleepnet,kiyasseh2021clocs}.

While it is challenging to accurately predict the next sleep stage, sleep generally follows typical patterns~\cite{feinberg1974changes}. It is unlikely that the sleep stage changes at every epoch.
When observing sleep patterns, short-term inertia or predictability appears. Additionally, the sleep pattern typically includes four to six long-term cycles overnight, each of which comprises both $REM$ and non-$REM$ stages~\cite{babloyantz1985evidence,crivello2019meaning}.

Our study proposes a sleep model that assesses and exploits the sequential nature of sleep. 
The use of sleep models can help to improve sleep classification accuracy, especially when relying on only simple surrogate sensors.
The sleep model concept originated from the language model that predicts the probability of the next word or character in speech or text. 
In particular, language models are important for improving the accuracy of speech recognition~\cite{hwang2016character,nakatani2019improving}. 

In Section~\ref{sec:related_works} we cover the background and related works, including the sleep data and signal processing models used in the study, as well as an introduction to language models. Section~\ref{sec:slm} presents the proposed sleep models and beam search decoding, while Section~\ref{sec:exp} provides experimental results and analyses.
Finally, Section~\ref{sec:conclusion} concludes the study.

%% file: 2.RelatedWorks.tex
\section{SLEEP DATA and BACKGROUND INFORMATION}\label{sec:related_works}
\subsection{Sleep Dataset}\label{sec:dataset}
There are many sleep datasets publically available due to a considerable number of PSG tests. These sleep datasets serve two purposes: firstly, to extract sleep sequences from simplified input signals using signal processing and deep neural networks; secondly, to develop sleep models based on sleep patterns observed in PSG tests.
To this end, we use the Haagleanden Medisch Centrum Sleep Database (HMC) dataset for the first purpose, and both HMC and NCH Sleep DataBank (NCHSDB) datasets for the second purpose~\cite{HMCdataset,zhang2018national}.
The HMC dataset was divided into 98 training, 24 validation, and 29 test splits, while the NCHSDB dataset used $3036$, $379$, and $380$ records for training, validation, and testing, respectively.
The signal data were pre-filtered and resampled at 100 Hz using the method described in~\cite{HMCdataset}.
Each epoch of the time-series data was classified into one of the five sleep stages ($W$, $REM$, $N1$, $N2$, and $N3$) based on the annotations provided for a sleep recording typically eight hours long.

\subsection{Signal Model}\label{sec:signal_model}
Our study involved the development of sleep signal-processing models capable of classifying sleep stages using either the entire PSG data or a subset of it.
The aim was to evaluate the performance of these models in both beam search decoding and greedy decoding.
In conventional automatic sleep-stage classification methods, the sleep class with the highest probability at each epoch is selected, a process referred to as greedy decoding in this context.
\input{figures/signal_model.tex}
To build our sleep signal model, we followed the DNN model proposed in~\cite{choi2022performance}, which consisted of feature extraction and classification blocks as depicted in Fig.~\ref{fig:signal_model}.
The input sensor data which included two EEG channels (C4-M1 and C3-M2), one EMG (chin), and one EOG (E2-E1) channel were applied to the feature extraction block.
The feature extraction block consisted of 3 convolutional neural networks (CNN) and 1 fully-connected DNN (FC-DNN)~\cite{lecun2015deep,lecun1995convolutional}.
In this block, the $e^{th}$ epoch that corresponded to 30-second of PSG data, $d_e$, was transformed to a $64$-dimensional feature vector, $f_e$.
The classification block employed in this study was formed using FC-DNN with a layer size of 320 and a depth of 2.
To classify one sleep stage, $s_e$, the classification block processed five epochs of input feature vectors: $f_{e-2}$, $f_{e-1}$, $f_e$, $f_{e+1}$, and $f_{e+2}$.
The HMC sleep dataset was used to develop two sleep signal models. The first model utilized four channels, which included EEG, EOG, and EMG as inputs.
The second model only employed EOG and had a single channel.

\subsection{Language Model}\label{sec:lm}
A language model predicts the probability distribution of the next word in a corpus. The inspiration for the proposed sleep model came from language models that are frequently used in computational linguistics and probabilistic fields~\cite{bengio2000neural,brown1992class}. An excellent example of such a large language model is ChatGPT, which is capable of generating human-like responses~\cite{brown2020language}.
They are traditionally trained on large corpus of text data, such as Wikipedia or speech transcriptions~\cite{lau2017grammaticality,bellegarda2004statistical}.
Traditionally, $n$-gram based language models have been used, but in recent years, DNNs such as LSTM-RNNs or Transformers have gained popularity due to their superior performance~\cite{hochreiter1997long,choi2018character,merity2017regularizing,vaswani2017attention,devlin2018bert}.
However, $n$-gram-based models are easier to build and require less time for inference~\cite{sundermeyer2012lstm,onan2021term,shareghi2019show}.

%% file: figures/signal_model.tex
\begin{figure}[b]
	\centering
	\includegraphics[width=0.7\columnwidth]{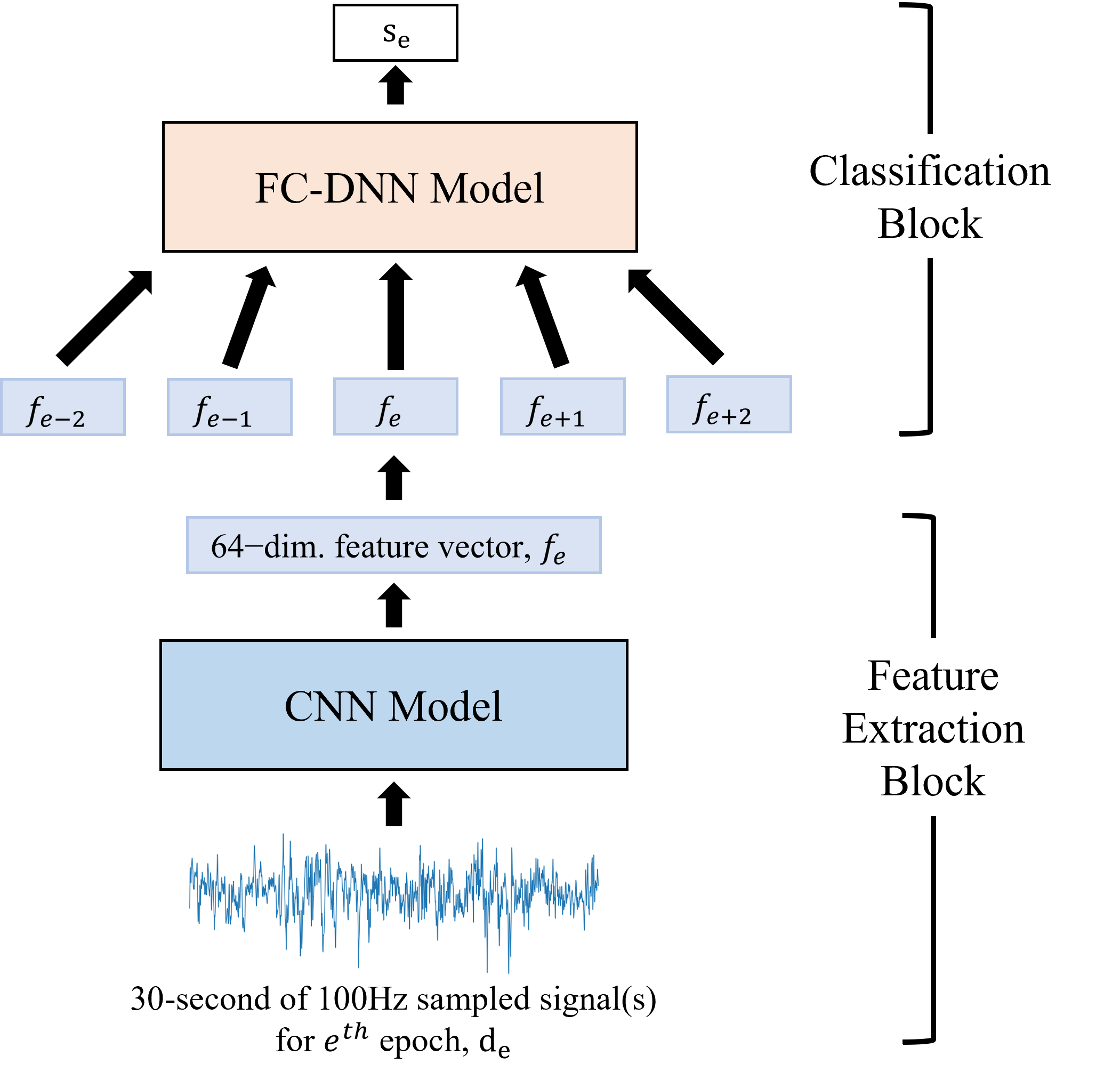}
	\caption{Signal model employing deep neural networks.}\label{fig:signal_model}
\end{figure}

%% file: 3.SleepModel_BeamSearchDecoding.tex
\section{Sleep Model and Beam Search Decoding}\label{sec:slm}
\subsection{Sleep Model}
The \textbf{SL}eep \textbf{M}odel (SLM) proposed in this study predicts the likelihood of each sleep class for the next epoch. These classes correspond to one of the five sleep stages, as shown in Fig.~\ref{fig:sleep_model}.
The SLM was constructed using training data that contains sleep sequences. Sleep stages exhibit fairly consistent patterns, and the following sleep class is heavily influenced by the preceding stages, much like language models~\cite{babloyantz1985evidence,feinberg1974changes}.

\input{figures/sleepModel.tex}

We built $n$-gram SLMs in manner of an $n$-gram language model.
An $n$-gram SLM was built in this study by approximating the probability with the number of occurrences of certain patterns in the training dataset. The number of occurrences was counted while traversing the entire sleep-stage sequences in dataset.
For example, a $3$-gram $\left(trigram\right)$ SLM shows the probability of the next sleep class based on the previous two sleep classes, which can have $25$ ($=5^2$) cases.
The $trigram$ probability of $P\left(\frac{N3}{N1, N2}\right)$, which is the probability of next stage being $N3$ following previous $N1$ and $N2$ stages, can be approximated as follows:
\begin{equation}
	\label{eq:ngram-prob.}
	P\left(\frac{N3}{N1, N2}\right) \approx \frac{C\left(N1, N2, N3\right)}{C\left(N1, N2\right)} \text{,}
\end{equation}
where $C\left(N1, N2\right)$ denotes the number of occurrences or counts of sleep stages $N1$ followed by $N2$ in the data.
Thus, the process of building an $n$-gram model is straightforward and the accuracy of prediction generally improves with the increasing value of $n$.
However, the above approximation cannot be accurate when the number of counts in the denominator, for example $C\left(N1, N2\right)$ in Equation (\ref{eq:ngram-prob.}), is not sufficiently large, which is called the data scarcity problem. To avoid the problem of data scarcity, a large training data is required.
Increasing $n$ also results in an exponential growth in the number of possible cases ($=5^{n-1}$). Generally, when the number of counts is very small, the modeling fallbacks to a lower $n$~\cite{chen1999empirical}.
In addition, there should not be a zero-probability prediction for the next class. Thus, when the numerator count is zero, we need smoothing that adds small numbers to the numerator and denominator terms to ensure a computational safety.
A length of 8 hours sleep consists of $960$ epochs or sleep stages, thus the total number of epochs for the HMC training set was approximately $100,000$.
To obtain a fairly accurate probability estimation, the number of the denominator count in Equation (\ref{eq:ngram-prob.}) needs to be around $100$.
If the sleep sequences are equally distributed, which is unrealistically optimistic, the number of different sequences that can be formed with $100,000$ epochs is about $1,000$, which can be approximated to $625\left(=5^4\right)$. Thus, we consider that the meaningful $n$-gram size for HMC dataset would be around $n = 5$.

We also developed SLMs based on the LSTM-RNN architecture by varying the number of LSTM layers or the hidden dimension of each LSTM layer~\cite{hochreiter1997long}.
The LSTM is a type of RNN architecture that has internal states to retain latent information from previous sequences.
This property makes LSTM based SLMs unrestricted by previous sequence length, in contrast to $n$-gram models, which are limited to a length of $n-1$.
The network architecture of the LSTM based SLM consists of a sleep stage embedding layer, LSTM layers, and a softmax output layer.

\subsection{Beam Search Decoding}
Typical automatic sleep-stage classification only employs a sleep signal model.
In the signal model, we can simply select the sleep class with the highest probability from each position in the sequence, which is often called greedy decoding.
When surrogate sensors are used for sleep tests, the accuracy of the signal model with greedy decoding is not sufficiently high.

The recognition accuracy of greedy decoding can be improved using a SLM.\@
The output of the SLM provides the prior information for sleep classification. The posterior probability was determined by multiplying, addition in the $\log$ domain, the probability of the signal model ($P_{sig}$) by that of the SLM ($P_{SLM}$) using Equation (\ref{eq:logprob_beam}).
\begin{equation}
	\label{eq:logprob_beam}
	\begin{split}
		\log P\left(s_e\right)=& 
		\log P_{sig}\left(s_e|d_{e-2}, \ldots, d_{e+2}\right) \\
		&+\alpha\log P_{SLM}\left(s_e|s_1, \ldots, s_{e-1}\right) \text{,}
	\end{split}
\end{equation}
where $d$ is input signal, $s$ denotes the sleep stages, $e$ in subscript means the $e^{th}$ epoch, and $\alpha$ is a parameter that assigns a balanced weight between the signal and sleep models.
Here, the signal model generates the likelihood of the sleep class using the input signal, whereas the SLM provides the prior probability.
Sleep-stage classification can be considered a sequence recognition problem, that is, we need to consider the results over a long time span.
The beam-search-decoding algorithm can select multiple sleep classes for each epoch in a given sequence~\cite{steinbiss1994improvements,ow1988filtered,tillmann2003word}.
This means that even sleep classes that do not show the highest probability can be saved for further evaluations in the future.
Because the number of sequence candidates grows exponentially as decoding progresses, it is not possible to retain all of them. The algorithm chooses the $W$-best alternatives via a hyperparameter known as the beam width.
The posterior probability for each epoch was obtained by multiplying the probabilities obtained from the signal and sleep models.
To determine the most likely beam, the posterior probabilities of each beam sequence are multiplied, and the resulting probability values are compared to select the highest one.

\subsection{Model Details and Metric}\label{sec:detail}
We built the $n$-gram model using KenLM, which implements fall-back, smoothing, and data compression~\cite{heafield2011kenlm}. We evaluated $n$-gram models on HMC and NCHSDB dataset, varying $n$ from $2$ to $9$, and applied the fall-back option. We also developed four kinds of LSTM-RNN based SLMs, 2 or 4 LSTM layers with 256 or 1,024 hidden dimensions on training split of each dataset.\@

The performance of language models is commonly measured by the perplexity~\cite{jelinek1977perplexity}.
The perplexity is defined as the inverse probability of the sequence of words, $w$, normalized by the number of words, $N$, as shown in Equation~\ref{eq:ppl},
\begin{equation}
	\label{eq:ppl}
	Perplexity\left(w_1, w_2, \ldots, w_N\right) = \sqrt[N]{\frac{1}{P\left(w_1, w_2, \ldots, w_N\right)}} \text{\phantom{:}.}
\end{equation}
We also used the perplexity to measure the performance of SLMs.
A lower perplexity value indicates better performance of the SLM in predicting the next sleep stage.

%% file: figures/sleepModel.tex
\begin{figure}[t]
	\centering
	\includegraphics[width=0.94\columnwidth]{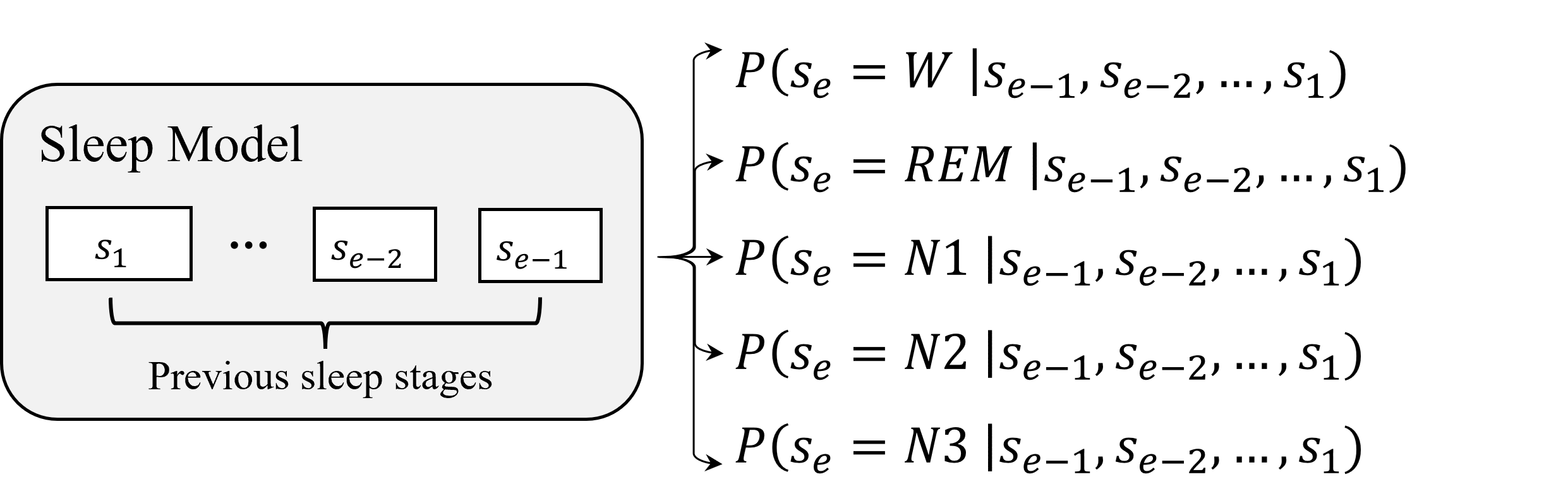}
	\caption{The sleep model that predicts the probabilities of sleep classes in the next epoch based on the previous ones.
	}\label{fig:slm}\label{fig:sleep_model}
\end{figure}

%% file: 4.ExperimentalResults.tex
\section{Experimental Results}\label{sec:exp}
\subsection{$n$-gram and LSTM-RNN based Sleep Models}\label{sec:exp:slm}
\input{tables/bigram_prob.tex}
\input{figures/fig_ngram.tex}

Table~\ref{tab:bigram_prob} shows the $2$-gram ($bigram$) probabilities on the HMC dataset. This table shows that the probability of $REM$ stage at the next epoch is 8\% when the current sleep stage is $N1$. As expected, the diagonal terms show high values, indicating the inertia of repeating the same sleep classes.
The transition probability from $N1$ to $N2$ is quite high and exhibits a low inertia of the $N1$ stage.
As a result, classification between $N1$ and $N2$ introduces many errors.

The perplexity of $n$-gram SLMs, when assessed with the test dataset, is shown in Fig.~\ref{fig:ngram_ppl}.
For HMC dataset, it shows that the perplexity decreased as $n$ increased until $n = 5$, but then increased thereafter.
As described in Section~\ref{sec:slm}, the back-off and smoothing algorithms implemented in KenLM have an impact on performance distortion when $n$ is greater than 5.
The size of NCHSDB dataset was approximately 20 times larger than HMC dataset, we can expect improved SLM performance and less sensitive to $n$.

Next we trained LSTM-RNN based SLMs using HMC or NCHSDB training set. Table~\ref{tab:hmc-nchsdb-lstm-ppl} shows the perplexity according to different LSTM-RNN based SLM configurations.
The results confirm that LSTM-RNN based SLMs are better than the $n$-gram based ones.

The experimental resutls indicated that the size of training set had a significant impact on the SLM performance.
The SLMs trained on the NCHSDB dataset performed relatively better than those with HMC.\@ This implies that building a good SLM requires at least $1,000$ records.
The performance of the SLM was affected by the characteristics of the sleep sequence, as evidenced by the difference in performance between the LSTM-RNN based SLM trained on the NCHSDB and tested on the HMC test set.
The SLM was configured with 2 LSTM layers and a hidden dimension of $1,024$, and achieved a perplexity of $1.286$ on the NCHSDB test set, but only $1.654$ on the HMC data. This difference is likely due to the fact that the sleep records in the HMC dataset are from adults, while those in the NCHSDB are from pediatrics, resulting in different sleep patterns that affect the inter-dataset performance of the SLMs.
\input{tables/lstm-ppl.tex}

\subsection{Beam Search Decoding for Combining Signal and Sleep Models}
We combined the 4-channel and 1-channel sleep signal models, explained in Section~\ref{sec:signal_model}, with the LSTM-RNN based SLM through beam-search decoding.
The 4-channel signal model was combined with the LSTM-RNN based SLM used in~\ref{sec:exp:slm} with the optimum $\alpha$, as described in Equation (\ref{eq:logprob_beam}), of $0.12$ and a beam width of $128$. The evaluation results are shown in Table~\ref{tab:beam-4ch}.
The classification results based on greedy decoding for the signal models that do not utilize SLMs are labeled as `Signal Model'.
Despite our attempts to improve decoding performance by incorporating the SLM, the results were not notably affected by varying $\alpha$ values ranging from $0.2$ to $1.5$.
This is likely due to the highly informative nature of the 4-channel signal model, which employed two channels of EEG and one channel each of EMG and EOG as input.

\input{tables/beamsearch-multi_table.tex}
\input{tables/alpha_change.tex}

The results in Table~\ref{tab:beam-1ch} demonstrate the 1-channel signal model when combined with the same SLM.\@
The combination resulted in a $6.5\%$ improvement in Kappa score and a $4.3\%$ improvement in accuracy, despite the lower accuracy of sleep-stage classification using EOG alone compared to the 4-channel PSG signal.
We used a beam width of 128 and searched for the best $\alpha$ on the validation set, and the results of the $\alpha$ search are listed in Table~\ref{tab:alpha-change}.

The findings indicate that utilizing SLMs in sleep-stage classification can be beneficial, particularly when working with limited signal information, such as a single input channel or surrogate signals. This approach has the potential to be applied to other input modalities or different contexts to improve sleep-stage classification.

%% file: tables/bigram_prob.tex
\begin{table}[t]
\caption{$2$-gram ($bigram$) sleep model probability table for HMC train split.
}\label{tab:bigram_prob}
\centering
\begin{adjustbox}{width=0.9\columnwidth}
{\setlength{\aboverulesep}{0.1em}
\setlength{\belowrulesep}{0.15em}
\footnotesize
\begin{tabular}{Scrrrrr}
\toprule
\multirow{2}{*}{\shortstack{\\Previous \\ Sleep Stage}}& 
\multicolumn{5}{Sc}{Next Sleep stage} \\
 \cmidrule(lr){2-6}
& $W$ & $REM$ & $N1$ & $N2$ & $N3$ \\
\midrule
$W$   & 0.854 & 0.001 & 0.138 & 0.003 & 0.000 \\
$REM$ & 0.016 & 0.907 & 0.066 & 0.010 & 0.000 \\
$N1$  & 0.109 & 0.080 & 0.498 & 0.311 & 0.000 \\
$N2$  & 0.019 & 0.014 & 0.062 & 0.864 & 0.040 \\
$N3$  & 0.007 & 0.001 & 0.007 & 0.063 & 0.921 \\
\bottomrule
\end{tabular}}
\end{adjustbox}
\end{table}

%% file: figures/fig_ngram.tex
\begin{figure}[b]
	\centering
	\includegraphics[width=0.6\columnwidth]
	{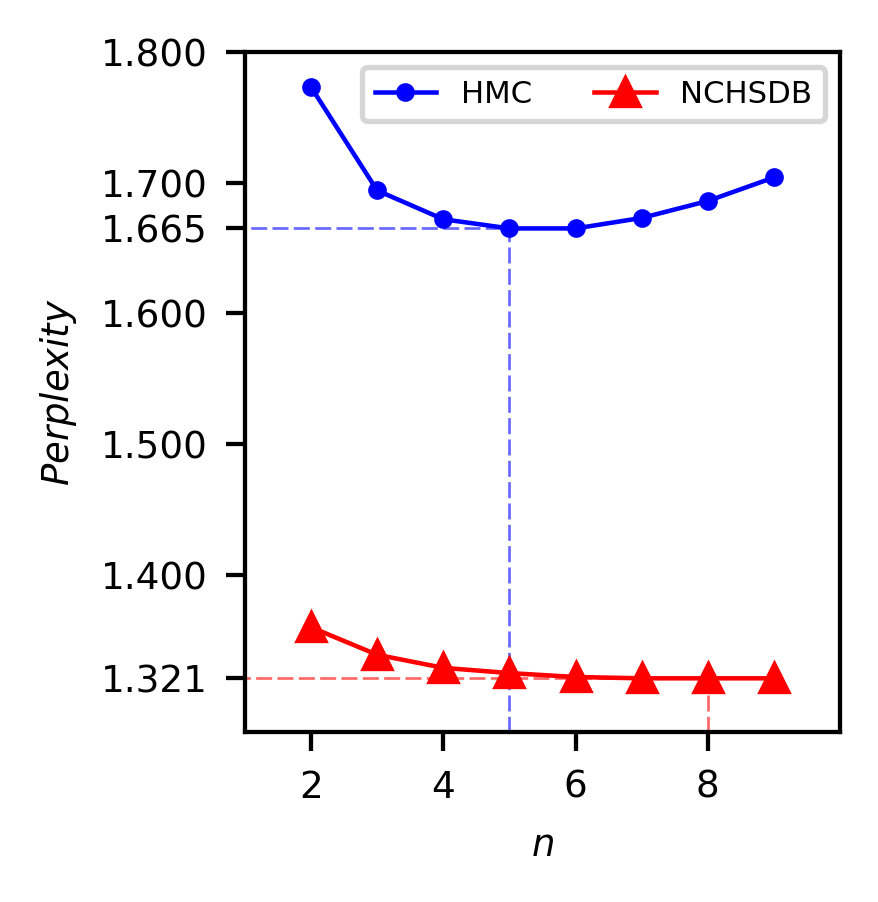}
	\caption{The test set perplexity of $n$-gram sleep models with HMC or NCHSDB dataset.}\label{fig:ngram_ppl}
\end{figure}

%% file: tables/lstm-ppl.tex
\begin{table}[t]
	\caption{The validation and test set perplexities of LSTM-RNN based SLMs trained with training split of HMC or NCHSDB dataset. The numbers in the first column represent (the number of layers) $\times$ (the size of the hidden dimension) for each SLM.\@}\label{tab:hmc-nchsdb-lstm-ppl}
	\centering
	\begin{adjustbox}{width=0.9\columnwidth}
		{\setlength{\aboverulesep}{0.1em}
		 \setlength{\belowrulesep}{0.15em}
			\footnotesize
			\begin{tabular}{Scrrrr}
				\toprule
				\multirow{2}{*}{\shortstack{\\LSTM-RNN \\ Sleep Model}} &
				\multicolumn{2}{Sc}{HMC} & \multicolumn{2}{Sc}{NCHSDB}\\
				\cmidrule(lr){2-3} \cmidrule(lr){4-5}
				                         & Valid.\ set                 & Test set       & Valid.\ set    & Test set \\
				\midrule
				2$\times$\phantom{0}256  &
				\textbf{1.546}           & 1.609                       & 1.293          & 1.287                     \\
				2$\times$1024            &
				1.547                    & \textbf{1.608}              & \textbf{1.291} & \textbf{1.286}            \\
				4$\times$\phantom{0}256  &
				1.547                    & 1.613                       & 1.293          & 1.287                     \\
				4$\times$1024            &
				1.548                    & 1.610                       & 1.291          & 1.286                     \\
				\bottomrule
			\end{tabular}}
	\end{adjustbox}
\end{table}

%% file: tables/beamsearch-multi_table.tex
\begin{table}[t]
	\caption{The performance of the LSTM-RNN SLMs with 4 or 1 -channel signal models. The SLMs had 2 layers of LSTM with 1,024 hidden dimensions.}
	\begin{subtable}{0.9\columnwidth}
		\caption{4-Channel Signal Model}\label{tab:beam-4ch}
		\centering
		\setlength{\aboverulesep}{0.1em}
		\setlength{\belowrulesep}{0.15em}
		\footnotesize
		\begin{tabular}{Sccrrrr}
			\toprule
			\multicolumn{1}{Sc}{\multirow{2}{*}{\shortstack{Sleep Stage                              \\Classification Model}}} &
			\multicolumn{1}{Sc}{\multirow{2}{*}{$\alpha$}}    &
			\multicolumn{2}{Sc}{\multirow{1}{*}{Valid.\ set}} &
			\multicolumn{2}{Sc}{\multirow{1}{*}{Test set}}                                           \\
			\cmidrule(lr){3-4}
			\cmidrule(lr){5-6}
			                                                  &       & Kappa & Acc.  & Kappa & Acc. \\
			\cmidrule(){1-6}
			Signal Model                                      &
			N/A                                               & 0.741 & 0.804 & 0.680 & 0.759        \\
			+ SLM Decording                                   &
			0.12                                              & 0.680 & 0.759 & 0.680 & 0.759        \\
			\bottomrule                                                                              \\
		\end{tabular}
	\end{subtable}
	\\\medskip 
	\begin{subtable}{0.9\columnwidth}
		\caption{1-Channel Signal Model}\label{tab:beam-1ch}
		\centering
		\setlength{\aboverulesep}{0.1em}
		\setlength{\belowrulesep}{0.15em}
		\footnotesize
		\begin{tabular}{Sccrrrr}
			\toprule
			\multicolumn{1}{Sc}{\multirow{2}{*}{\shortstack{Sleep Stage                              \\Classification Model}}} &
			\multicolumn{1}{Sc}{\multirow{2}{*}{$\alpha$}}    &
			\multicolumn{2}{Sc}{\multirow{1}{*}{Valid.\ set}} &
			\multicolumn{2}{Sc}{\multirow{1}{*}{Test set}}                                           \\
			\cmidrule(lr){3-4}
			\cmidrule(lr){5-6}
			                                                  &       & Kappa & Acc.  & Kappa & Acc. \\
			\cmidrule(){1-6}
			Signal Model                                      &
			N/A                                               & 0.505 & 0.629 & 0.464 & 0.602        \\
			+ SLM Decording                                   &
			0.42                                              & 0.596 & 0.698 & 0.519 & 0.644        \\
			\bottomrule
		\end{tabular}
	\end{subtable}
\end{table}

%% file: tables/alpha_change.tex
\begin{table}[t]
\caption{The effect of probability weighting factor. The beam width was 128 and 1-channel signal model with 2$\times$1024 LSTM-RNN SLM was evaluated.}\label{tab:alpha-change}
\centering
\begin{adjustbox}{width=0.75\columnwidth}
{\setlength{\aboverulesep}{0.25pt}
\setlength{\belowrulesep}{0.25pt}
\footnotesize
\begin{tabular}{Scrrrr}
\toprule
\multicolumn{1}{Sc}{\multirow{2}{*}{$\alpha$}} & 
\multicolumn{2}{Sc}{\multirow{1}{*}{Valid.\ set}} & 
\multicolumn{2}{Sc}{\multirow{1}{*}{Test set}} \\
\cmidrule(lr){2-3} \cmidrule(lr){4-5}
 & Kappa & Acc. & Kappa & Acc. \\
 \midrule
 0.34 &  0.587 & 0.692 & 0.517 & 0.642 \\
 0.38 &  0.592 & 0.696 & 0.516 & 0.642 \\
 0.42 &  \textbf{0.596} & \textbf{0.698} & \textbf{0.519} & \textbf{0.644} \\
 0.46 &  0.588 & 0.693 & 0.515 & 0.640 \\
 0.50 &  0.594 & 0.697 & 0.511 & 0.638 \\
\bottomrule
\end{tabular}}
\end{adjustbox}
\end{table}

%% file: 5.ConculdingRemark.tex
\section{Conclusion}\label{sec:conclusion}
Our study proposed sleep models for predicting the next sleep stage, which can significantly enhance sleep classification accuracy by utilizing information from numerous sleep-stage sequences and extending the context length.
Our models were applied to sleep-stage classification using simple EOG sensors.
Sleep-stage sequences, rather than the signal sources, are the only data required to train sleep models, making them compatible with various sleep archives.
Furthermore, once sleep models have been built, they can be employed in sleep-stage classification using different sensors, such as PPG, EOG, or ECG.\@